\documentclass[preprint,showpacs,aps,superscriptaddress]{revtex4}   
\usepackage{graphicx}
\usepackage{bm}
\usepackage[utf8]{inputenc}
\usepackage{amsmath}
\usepackage{color}

\newcommand{\UQ}{School of Mathematics and Physics, University of Queensland, Brisbane, 
QLD 4072, Australia.}

\newcommand{\Griffith}{Centre for Quantum Dynamics, Griffith University, Gold Coast, QLD 4215, Australia}

\newcommand{\etal}{{\em et al.}}

\newcommand{\pr}{Phys. Rev. }
\newcommand{\jpb}{J. Phys. B }

\definecolor{maroon}{rgb}{0.7,0,0}

\definecolor{ngreen}{rgb}{0.3,0.7,0.3}

\definecolor{golden}{rgb}{0.8,0.6,0.1}

\begin{document}
\title{The versatility of continuous-variable asymmetric tripartite entanglement allows Alice and Clare to keep secrets from Bob}

\author{M.~K. Olsen}
\affiliation{\UQ}
\author{E.~G. Cavalcanti}
\affiliation{\Griffith}
\date{\today}

\begin{abstract}

The fully symmetric Gaussian tripartite entangled pure states will not exhibit two-mode Einstein Podolsky-Rosen (EPR)-steering. This means that any two participants cannot share quantum secrets using the security of one-sided device independent quantum key distribution (1SDI-QKD) without involving the third. They are restricted at most to standard quantum key distribution (S-QKD), which is less secure. Here we demonstrate an asymmetric tripartite system that can exhibit bipartite EPR-steering, so that two of the participants can use 1SDI-QKD without involving the other. This is possible because the promiscuity relations of continuous-variable tripartite entanglement are different from those of discrete-variable systems. We analyse these properties for two different systems, showing that the asymmetric system exhibits practical properties not found in the symmetric one.

\end{abstract}


\maketitle

\section{Introduction}
\label{sec:intro}

Quantum key distribution (QKD) is the first mature technology which uses fundamental quantum mechanics~\cite{QKDScarani}, and allows for the creation of a secret key between
authorised partners connected by a quantum channel and a classical authenticated channel. The field began with the presentation
of the first complete protocol by Bennett and Brassard~\cite{BB84}, based on ideas developed by Wiesner~\cite{Wiesner}. QKD can be performed with both discrete and continuous-variable systems and can be divided into three basic categories, with different security categorisations~\cite{Cyril} and different degrees of quantum correlations needed to function effectively. The first of these is standard QKD (S-QKD), where both Alice and Bob trust their preparation and measurement devices, which requires only entanglement from the hierarchy defined by Wiseman \etal~\cite{Wisesteer}. The second, known as one-sided device independent QKD (1SDI-QKD), only requires that one apparatus be trusted, and requires correlations at the level of Einstein-Podolsky-Rosen (EPR)~\cite{EPR} steering. The third, known as device independent QKD (DI-QKD) requires correlations at the level of Bell violations~\cite{Bell}. Here we analyse a feasible continuous-variable regime in which 1SDI-QKD is available between two participants in a fully tripartite entangled system, while the third cannot participate. Calling the participants Alice, Bob, and Clare, we show that for some systems Alice and Clare can exchange secret keys which are not accessible to Bob. We will also demonstrate that this is not possible in fully symmetric tripartite entangled systems.

Adesso \etal~\cite{Adesso1} have provided a classification scheme of continuous-variable Gaussian three-mode states with five distinct classes. The first classification describes states which are not separable under any of the three possible bipartitions. A subset of these states, which are invariant under the exchange of any two modes, are known as fully symmetric. Two examples of these are the states analysed by Aoki \etal~\cite{Aoki}, which mixes three squeezed states on two beamsplitters, and the triply concurrent downconversion scheme analysed by Smithers \etal~\cite{Smithers} and Bradley \etal~\cite{Pfister}. The fully symmetric states as defined by Adesso \etal~\cite{Adesso1} have the property that their covariance matrices can all be put in the same form by transformations which do not change the entanglement properties~\cite{Serafini2005,Adesso2004}, and Gaussian states have the property that they are completely characterised by the covariance matrix. This means that proving general properties of the fully symmetric states can be relatively simple.  
There are other possible tripartite Gaussian entangled states, one example of which is produced by a scheme which combines downconversion and sum frequency generation~\cite{Fortescue,Ferraro,Yu,OlsenBradley,Clervie}, and does not possess the symmetry of the fully symmetric states. We will call this class of states asymmetric. Necessarily, these states will not have covariance matrices that can be put into an identical standard form. Any given asymmetric state will have a unique covariance matrix. This makes general proofs of the properties of asymmetric states more difficult, but as we merely wish to prove that something is possible, showing it for one member of the class of states is sufficient. We will show, that for the example of a fully tripartite entangled state considered here, 1SDI-QKD between two of the participants is indeed possible. As a demonstration of principle, we begin by examining a simple travelling wave model. We then analyse the appropriate correlation functions of the asymmetric scheme in the more experimentally relevant situation where the nonlinear medium is contained inside a pumped optical cavity.

\section{Entanglement and EPR inequalities}
\label{sec:inequalities}

With the annihilation operator $\hat{a}_{i}$ corresponding to mode $i$, we define the quadrature operators as $\hat{X}_{i}=\hat{a}_{i}+\hat{a}^{\dag}_{i}$ and $\hat{Y}_{i}=-i(\hat{a}_{i}-\hat{a}^{\dag}_{i})$.
The bipartite Einstein-Podolsky-Rosen (EPR) paradox~\cite{EPR} is detected by the well-known criteria developed by Reid~\cite{eprMDR}, in terms of inferred quadrature variances,
\begin{equation}
\Pi V_{ij} = V^{inf}(\hat{X}_{i})V^{inf}(\hat{Y}_{i})\geq 1,
\label{eq:eprMDR}
\end{equation}
with an EPR state being indicated by violation of the inequality.
This condition is optimal for bipartite Gaussian systems. Since this criterion was developed, Wiseman \etal~\cite{Wisesteer} have formalised the concept of EPR-steering mathematically, showing that it is equivalent to the informal concept of steering developed by Schrödinger~\cite{Erwin}. We will therefore refer to states which satisfy the criterion as possessing EPR-steering. In the above, $\Pi V_{ij}$ is the product of the inferred variances of mode $i$, as inferred by the operator with access to mode $j$. For pure Gaussian states and Gaussian measurements, such as those considered in the majority of this article, the Reid criterion is both necessary and sufficient to demonstrate EPR-steering~\cite{Jonesteer}.

Bipartite entanglement can be established in terms of the functions of quadrature variance inequalities developed by Duan and Simon~\cite{Duan,Simon}, 
\begin{equation}
V(\hat{X}_{i}\pm\hat{X}_{j})+V(\hat{Y}_{i}\mp\hat{Y}_{j}) \geq 4,
\label{eq:DSinequalities}
\end{equation}
with violation of either of these being a demonstration of bipartite entanglement. We will call the first of these $DS_{ij}^{+}$ and the second $DS_{ij}^{-}$. Because the travelling wave systems we consider are Gaussian and pure, these entanglement correlations are both necessary and sufficient for the demonstration of bipartite entanglement~\cite{Teh}.

We will establish tripartite entanglement using the van Loock-Furusawa conditions~\cite{VLF}, which give a set of three inequalities
\begin{equation}
V_{ij} = V(\hat{X}_{i}-\hat{X}_{j})+V(\hat{Y}_{i}+\hat{Y}_{j}+g_{k}\hat{Y}_{k}) \geq 4, 
\label{eq:VLF}
\end{equation}
for which the violation of any two demonstrates tripartite entanglement. The $g_{j}$, which are arbitrary and real, can be optimised~\cite{AxMuzz}, using the variances and covariances, as
\begin{equation}
g_{i} = -\frac{V(\hat{Y}_{i},\hat{Y}_{j})+V(\hat{Y}_{i},\hat{Y}_{k})}{V(\hat{Y}_{i})},
\label{eq:VLFopt}
\end{equation}
which is the process we follow here. Another set of inequalities was also presented by van Loock and Furusawa, the violation of any one of which is sufficient to prove tripartite entanglement,
\begin{equation}
V_{ijk} = V(\hat{X}_{i}-\frac{\hat{X}_{j}+\hat{X}_{k}}{\sqrt{2}})+V(\hat{Y}_{i}+\frac{\hat{Y}_{j}+\hat{Y}_{k}}{\sqrt{2}}) \geq 4.
\label{eq:VLFijk}
\end{equation}

\section{Symmetric systems}
\label{sec:symm}

\begin{figure}[tbp]
\includegraphics[width=0.6\columnwidth]{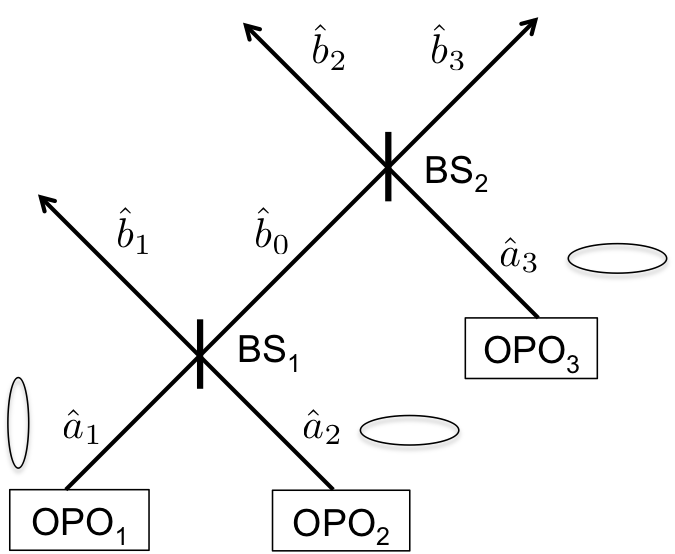}
\caption{ The beam-splitter model used for symmetric systems. Three optical parametric oscillators produce outputs squeezed in the $\hat{Y}$ ($\mathrm{OPO}_1$) or $\hat{X}$ ($\mathrm{OPO}_2$ and $\mathrm{OPO}_3$) quadratures. The outputs are combined with two beam-splitters, producing three output fields $\hat{b}_1, \hat{b}_2, \hat{b}_3$.}
\label{fig:BSmodel}
\end{figure}

As our fully symmetric system, we will use the beamsplitter model of Aoki \etal~\cite{Aoki}, shown in Fig.~\ref{fig:BSmodel}. One of the pioneering systems for continuous variable tripartite entanglement, this came from van Loock and Braunstein~\cite{VLB}, and was implemented experimentally by
Aoki \etal, who mixed three squeezed states on two beamsplitters to obtain three entangled output beams. This setup is a subset of the systems recently analysed by Wang \etal~\cite{WangGHZ}.
The system uses three optical parametric oscillators (OPO), with the first, OPO$_{1}$, producing a state squeezed in the $\hat{Y}$ quadrature, while the other two produce $\hat{X}$ squeezed states. The annihilation operator $\hat{a}_{j}$ represents the output of OPO$_{j}$. The output of OPO$_{1}$ and OPO$_{2}$ are mixed on the first beamsplitter, BS$_{1}$, to produce outputs represented by $\hat{b}_{0}$ and $\hat{b}_{1}$. The field corresponding to $\hat{b}_{0}$  is then mixed with $\hat{a}_{3}$ on BS$_{2}$. The outputs of BS$_{2}$ are represented by $\hat{b}_{2}$ and $\hat{b}_{3}$. With the squeezed inputs, tripartite entanglement is found between the three outputs.

 Assigning BS$_{1}$ a reflectivity of $\mu$ and BS$_{2}$ a reflectivity of $\nu$, with $\mu=2/3$ and $\nu=1/2$ as in Aoki \etal~\cite{Aoki} to give a fully symmetric state, we find the solutions for the $\hat{b}_{j}$ in terms of the inputs as 
 \begin{eqnarray}
 \hat{b}_{1} &=& \sqrt{1-\mu}\;\hat{a}_{1}+\sqrt{\mu}\;\hat{a}_{2}, \nonumber \\
 \hat{b}_{2} &=& \sqrt{\mu(1-\nu)}\;\hat{a}_{1}-\sqrt{(1-\mu)(1-\nu)}\;\hat{a}_{2}+\sqrt{\nu}\;\hat{a}_{3}, \nonumber \\
 \hat{b}_{3} &=& \sqrt{\mu\nu}\;\hat{a}_{1}-\sqrt{\nu(1-\mu)}\;\hat{a}_{2}-\sqrt{1-\nu}\;\hat{a}_{3},
\label{eq:BSErnie} 
\end{eqnarray}
which allow us to find all the correlations we require for the bipartite and tripartite correlations we wish to calculate. The required variances and covariances are given in Ref.~\cite{EPR3}, and are all that is necessary to calculate the Duan-Simon, van Loock-Furusawa and EPR-steering correlations.

For a squeezing parameter $r$, we may assume minimum uncertainty squeezed input states (for more realistic inputs, see Ref.~\cite{EPR3}) and set 
\begin{eqnarray}
V(\hat{X}_{a_{1}}) &=& V(\hat{Y}_{a_{2}}) = V(\hat{Y}_{a_{3}}) = \mbox{e}^{r}, \nonumber \\
V(\hat{Y}_{a_{1}}) &=& V(\hat{X}_{a_{2}}) = V(\hat{X}_{a_{3}}) = \mbox{e}^{-r},
\label{eq:minimumuncertainty}
\end{eqnarray}
which leads to the bipartite correlations of Eq.~(\ref{eq:DSinequalities}),
\begin{equation}
DS_{ij}^{\pm} = 4\cosh r \pm \frac{8}{3}\sinh r,
\label{eq:DSDS}
\end{equation}
of which the $DS_{ij}^{-}$ fall below $4$ over a range of $r$. The inferred variances are found as
\begin{equation}
V^{inf}(\hat{X}_{i})=\left[V^{inf}(\hat{Y}_{i}) \right]^{-1} = \frac{3\cosh r + \sinh r}{2+\mbox{e}^{2r}},
\label{eq:VinfBS}
\end{equation}
so that the Reid EPR correlation of  Eq.~(\ref{eq:eprMDR}) is equal to unity for all $r$. We can also see that two-mode EPR is not possible for this system following the approach taken by Wang \etal~\cite{WangGHZ}. Those authors considered the possibility of EPR-steering in the outputs of a cascading beamsplitter scheme of the type considered above, with $N$ OPOs, $N-1$ beamsplitters, and $N$ output modes, with the first OPO squeezed in the $\hat{X}$ quadrature and the remaining ones in the $\hat{Y}$ quadrature. If $M$ of the output modes are used to steer any other of the output modes, EPR-steering can be demonstrated when the following function falls below one,
\begin{equation}
E_{i|M} = \frac{2(M+1)(N-M-1)[\cosh 4r-1]+N^{2}}{2M(N-M)[\cosh 4r-1]+N^{2}}\;.
\label{eq:WangEPR}
\end{equation}
In the case we consider, with $N=3$ output modes it is easily seen that the above function is equal to $1$ for $N=3$ and $M=1$, meaning that EPR-steering is not possible between any two of the output modes of the system above, for any value of the input squeezing parameters. 

The optimised $V_{ij}$ of Eq.~\eqref{eq:VLF} are found as
\begin{equation}
V_{ij} = \frac{2+10\mbox{e}^{2r}}{\mbox{e}^{r}+2\mbox{e}^{3r}},
\label{eq:VijAoki}
\end{equation}
and the $V_{ijk}$ of Eq.~\eqref{eq:VLFijk} are
\begin{equation}
V_{ijk} = 4\left(\cosh r-\frac{2\sqrt{2}}{3}\sinh r \right),
\label{eq:VijkAoki}
\end{equation}
with these not changing under permutations of the indices. Note that, with optimisation, the $V_{ij}$ begin at $4$, rather than at the $5$ found in Ref.~\cite{EPR3} without optimisation. The uniqueness of the standard form of the covariance matrix for a pure fully symmetric Gaussian tripartite state, as shown by Adesso \etal~\cite{Adesso1}, means that all possible symmetric Gaussian three-mode systems have a covariance matrix which can be written in the same form. This means that Eq.~\eqref{eq:VinfBS} shows that bipartite EPR-steering is not possible in any fully symmetric Gaussian tripartite system. This result shows that at most S-QKD is possible between any pair of Alice, Bob, and Clare with this system. All three must participate in any device independent QKD. Although not shown here, the violation of three-mode EPR-steering correlations~\cite{EPR3} for this system shows that the participants can combine pairwise to perform 1SDI-QKD with the third participant~\cite{notGHZ}.

\section{An asymmetric system}
\label{sec:asym}

Our asymmetric system, which combines downconversion with sum-frequency generation, was first proposed by Smithers and  Lu~\cite{Smithers}, and theoretically analysed in a travelling wave configuration by Ferraro \etal~\cite{Ferraro}, and in an intracavity configuration by Yu \etal~\cite{Yu}. The configuration was subsequently analysed in more depth by Pennarun \etal~\cite{Clervie}, who investigated the stability properties and predicted tripartite entanglement in different regimes. It consists of a nonlinear medium pumped at frequency $\omega_{0}$. The downconversion part of the process, denoted by the effective nonlinearity $\kappa_{1}$, generates two fields at  
$\omega_{1}$ and $\omega_{3}$, where $\omega_{0}=\omega_{1}+\omega_{3}$. The pump field at $\omega_{0}$ can then combine with the field at $\omega_{3}$ in a sum frequency generation process~\cite{SFG}, to produce a further field at $\omega_{2}$, with effective nonlinearity $\kappa_{2}$. We will use the annihilation operators $\hat{a}_{j}$ to describe the fields at $\omega_{j}$ for $j=1,2,3$. If we consider that the pump field is intense and classical so that depletion does not become important, we may write the interaction Hamiltonian as
\begin{equation}\label{eq:Hint}
{\cal H}_{int} =
i\hbar\kappa_{1}(\hat{a}_1^\dag\hat{a}_3^\dag-\hat{a}_1\hat{a}_3)+i\hbar\kappa_{2}(\hat{a}_3\hat{a}_2^\dag-\hat{a}_3^\dag\hat{a}_2).
\end{equation}
In what follows we will define the variable $\zeta$ as $\kappa_{1}^{2}-\kappa_{2}^{2}$.

We find that the Heisenberg equations of motion for the annihilation and creation operators, being linear, may be solved analytically after Jordan decomposition of the time evolution matrix, as described in standard undergraduate textbooks~\cite{anta}. We used Mathematica for this, allowing us  
to write solutions for the quadrature operators (noting that $\kappa_{1}>\kappa_{2}$ here),
\begin{eqnarray}
\hat{X}_{1}(t) &=&  \frac{\kappa_{1}^{2}\cosh \zeta t-\kappa_{2}^{2}}{\zeta^{2}}\hat{X}_{1}(0)-\frac{\kappa_{1}\kappa_{2}(\cosh\zeta t-1)}{\zeta^{2}}\hat{X}_{2}(0)+\frac{\kappa_{1}\sinh\zeta t}{\zeta^{2}}\hat{X}_{3}(0), \nonumber \\
\hat{X}_{2}(t) &=&\frac{\kappa_{1}\kappa_{2}(\cosh\zeta t-1)}{\zeta^{2}}\hat{X}_{1}(0) + \frac{\kappa_{1}^{2}-\kappa_{2}^{2}\cosh\zeta t}{\zeta^{2}}\hat{X}_{2}(0)+\frac{\kappa_{1}\sinh\zeta t}{\zeta^{2}}\hat{X}_{3}(0), \nonumber \\
\hat{X}_{3}(t) &=& \frac{\kappa_{1}\sinh\zeta t}{\zeta^{2}}\hat{X}_{1}(0)-\frac{\kappa_{2}\sinh\zeta t}{\zeta^{2}}\hat{X}_{2}(0)+\cosh\zeta t\;\hat{X}_{3}(0), \nonumber \\
\hat{Y}_{1}(t) &=& \frac{\kappa_{1}^{2}\cosh\zeta t-\kappa_{2}^{2}}{\zeta^{2}}\hat{Y}_{1}(0)+\frac{\kappa_{1}\kappa_{2}(\cosh\zeta t-1)}{\zeta^{2}}\hat{Y}_{2}(0)-\frac{\kappa_{1}\sinh\zeta t}{\zeta^{2}}\hat{Y}_{3}, \nonumber \\
\hat{Y}_{2}(t) &=& -\frac{\kappa_{1}\kappa_{2}(\cosh\zeta t -1)}{\zeta^{2}}\hat{Y}_{1}(0)+\frac{\kappa_{1}^{2}-\kappa_{2}^{2}\cosh\zeta t}{\zeta^{2}}\hat{Y}_{2}(0)+\frac{\kappa_{1}\sinh\zeta t}{\zeta^{2}}\hat{Y}_{3}(0), \nonumber \\
\hat{Y}_{3}(t) &=& -\frac{\kappa_{1}\sinh\zeta t}{\zeta^{2}}\hat{Y}_{1}(0)-\frac{\kappa_{2}\sinh\zeta t}{\zeta^{2}}\hat{Y}_{2}(0)+\cosh\zeta t\;\hat{Y}_{3}(0),
\label{eq:CPXYsols}
\end{eqnarray}
which allows us to find expressions for all the entanglement and EPR-steering correlations. Setting
\begin{eqnarray}
\alpha &=& \frac{\kappa_{1}^{2}\cosh \zeta t-\kappa_{2}^{2}}{\zeta^{2}}, \nonumber \\
\beta &=& \frac{\kappa_{1}\kappa_{2}(\cosh\zeta t-1)}{\zeta^{2}}, \nonumber \\
\gamma &=& \frac{\kappa_{1}\sinh\zeta t}{\zeta^{2}}, \nonumber \\
\delta &=& \frac{\kappa_{1}^{2}-\kappa_{2}^{2}\cosh\zeta t}{\zeta^{2}}, \nonumber \\
\epsilon &=& \frac{\kappa_{2}\sinh\zeta t}{\zeta^{2}}, \nonumber \\
\eta &=& \cosh\zeta t,
\label{eq:Greek}
\end{eqnarray}
and noting that the single quadrature expectation values vanish when the input modes
are vacuum, we find the moments required for the variances and covariances as
\begin{eqnarray}
\langle \hat{X}_{1}^{2}\rangle &=& \langle \hat{Y}_{1}^{2}\rangle = \alpha^{2}+\beta^{2}+\gamma^{2}, \nonumber \\
\langle \hat{X}_{2}^{2}\rangle &=&  \langle \hat{Y}_{2}^{2}\rangle =  \beta^{2}+\delta^{2}+\gamma^{2}, \nonumber \\
\langle \hat{X}_{3}^{2}\rangle &=&  \langle \hat{Y}_{3}^{2}\rangle =  \gamma^{2}+\epsilon^{2}+\eta^{2}, \nonumber \\
\langle \hat{X}_{1}\hat{X}_{2}\rangle &=& \alpha\beta-\beta\delta+\gamma^{2}, \nonumber \\
\langle \hat{X}_{1}\hat{X}_{3}\rangle &=& \alpha\gamma+\beta\epsilon+\gamma\eta, \nonumber \\
\langle \hat{X}_{2}\hat{X}_{3}\rangle &=& \gamma\beta-\delta\epsilon+\gamma\eta, \nonumber \\
\langle \hat{Y}_{1}\hat{Y}_{2}\rangle &=& -\alpha\beta+\beta\delta-\gamma^{2}, \nonumber \\
\langle \hat{Y}_{1}\hat{Y}_{3}\rangle &=& -\alpha\gamma-\beta\epsilon-\gamma\eta, \nonumber \\
\langle \hat{Y}_{2}\hat{Y}_{3}\rangle &=& \beta\gamma-\delta\epsilon+\gamma\eta.
\label{eq:CPXYmoments}
\end{eqnarray}

\begin{figure}[tbp]
\includegraphics[width=0.75\columnwidth]{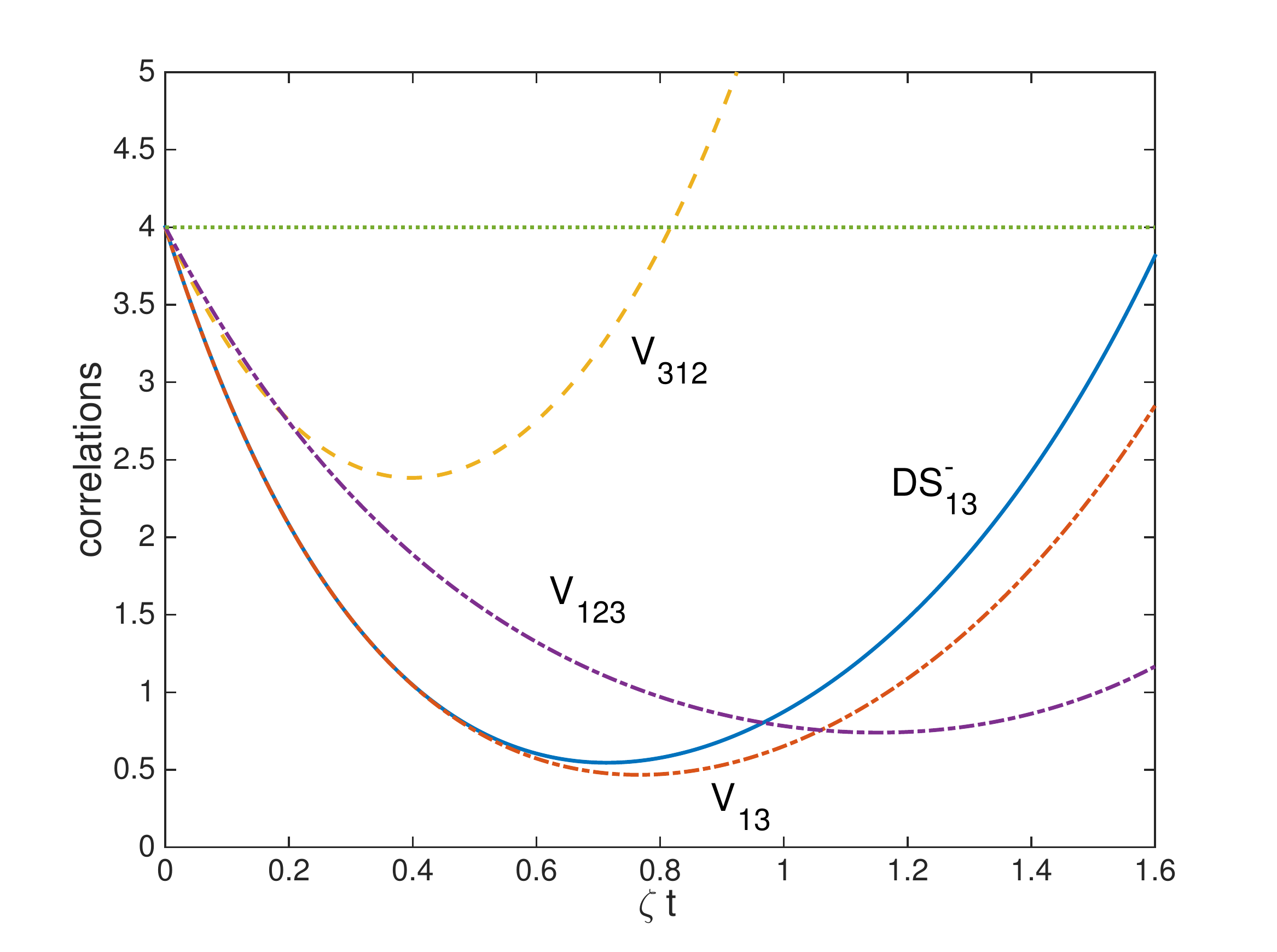}
\caption{(colour online) DS$_{13}^{-}$, V$_{123}$, V$_{312}$, and V$_{13}$ for the asymmetric model, with $\kappa_{2}=0.6\kappa_{1}$. We see that both bipartite and tripartite entanglement are predicted over a range of interaction strength $\zeta t$. Note that the line at $4$ is a guide to the eye.}
\label{fig:Clervie}
\end{figure}

\begin{figure}[tbp]
\includegraphics[width=0.75\columnwidth]{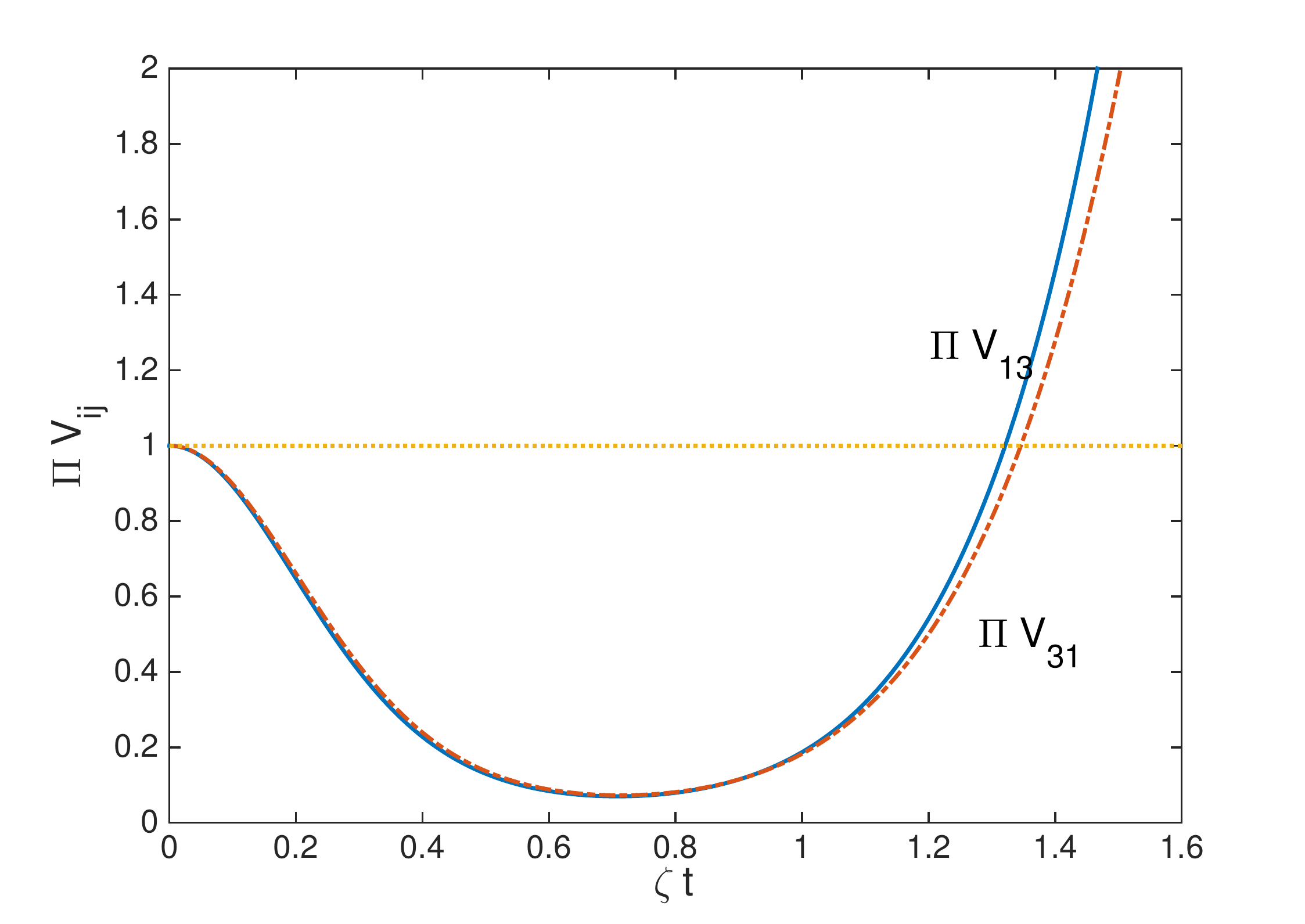}
\caption{(colour online) The EPR-steering correlations for modes $1$ and $3$ of the asymmetric model, with $\kappa_{2}=0.6\kappa_{1}$. Note that the line at $1$ is a guide to the eye.}
\label{fig:EPRClervie}
\end{figure}

Setting $\kappa_{2}=0.6\kappa_{1}$, we find that the correlation $V_{123}$ is the most sensitive for the detection of tripartite entanglement, as shown in Fig.~\ref{fig:Clervie}.  According to Eq.~\eqref{eq:VLFijk}, this correlation being less than four is sufficient to demonstrate that the system exhibits full tripartite entanglement. We find that only one of the possible bipartitions exhibits bipartite entanglement, with this being shown by DS$_{13}^{-}$ of Eq.~\eqref{eq:DSinequalities}. None of the other bipartitions were found to violate the necessary inequalities and have not been shown here. We also see, in Fig.~\ref{fig:EPRClervie}, that there is bipartite EPR-steering between modes $1$ and $3$, with the expressions needed for the correlations of Eq.~\eqref{eq:eprMDR} given by
\begin{eqnarray}
\Pi V_{13} &=& \frac{\left[(\alpha^{2}+\beta^{2}+\gamma^{2})(\gamma^{2}+\epsilon^{2}+\eta^{2})-(\alpha\gamma+\beta\epsilon+\gamma\eta)^{2}\right]^{2}}{(\gamma^{2}+\epsilon^{2}+\eta^{2})^{2}}, \nonumber\\
\Pi V_{31} &=& \frac{\left[(\alpha^{2}+\epsilon^{2}+\eta^{2})(\beta^{2}+\gamma^{2}+\delta^{2})-(\alpha\gamma+\beta\epsilon+\gamma\eta)^{2}\right]^{2}}{(\beta^{2}+\delta^{2}+\gamma^{2})^{2}},
\label{eq:symbolic}
\end{eqnarray}
where the regimes where these expressions fall below unity then best seen by plotting of the analytical results.
We find that no EPR-steering is possible between any of the other pairs. The EPR-steering between Alice and Clare is almost symmetric for the input parameters used here, although asymmetric EPR-steering~\cite{sapatona} can be seen for other parameters. As we have assigned modes to Alice, Bob, and Clare in numerical order, this means that there is a possibility that Alice and Clare can share secrets using 1SDI-QKD, without involving Bob. Bob cannot even participate in S-QKD with either of the other two. This feature is a direct result of the promiscuity of asymmetric continuous-variable tripartite entanglement, and is a demonstration of the flexibility of continuous-variable systems. When we examine~\cite{notGHZ} the tripartite EPR-steering correlations of Ref.~\cite{EPR3}, we find that Bob and Clare together can steer Alice, while Alice and Bob together only violate the necessary inequality marginally. Alice and Clare together cannot steer Bob at all for the parameters used here.

\begin{figure}[tbp]
\includegraphics[width=0.75\columnwidth]{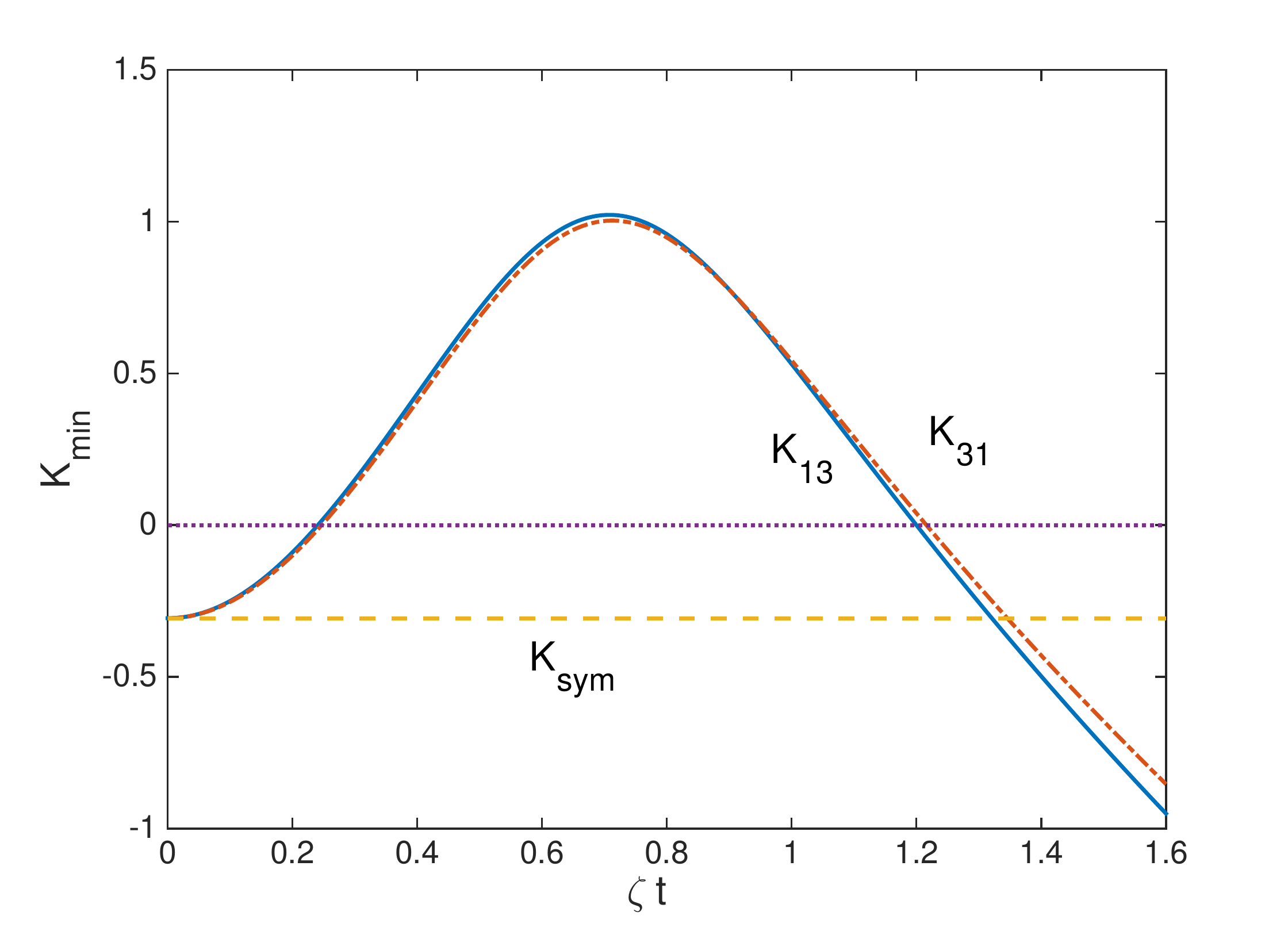}
\caption{(colour online) The minimum bit rates for modes $1$ and $3$ of the asymmetric model, with $\kappa_{2}=0.6\kappa_{1}$, and the minimum rate for any pair of the symmetric scheme. $K_{ij}$ is the minimum bit rate for mode $j$ sending a quantum key to mode $i$.  A positive value shows that 1SDI-QKD is possible. The line at zero is a guide to the eye.}
\label{fig:Kmin}
\end{figure}

To ensure that 1SDI-QKD is in fact viable between Alice and Clare, we also need to calculate the bit rate. As shown in \cite{Cyril}, EPR-steering is necessary, but not sufficient for 1SDI-QKD. The continuous-variable case was analysed by Walk \etal~\cite{Walk}. In their work, the secret key is encoded in the $\hat{X}$ quadrature of a field mode $j$, and a sufficiently strong demonstration of CV EPR-steering from mode $i$ to $j$ bounds the information that an eavesdropper can have about the key, in a way that is independent of the device at mode $i$, so that a secret key rate obtained using reverse reconciliation is lower bounded by 
\begin{equation}
K_{min} \geq \log\left(\frac{2\mbox{e}^{-1}}{\sqrt{\Pi V_{ij}}}\right).
\label{eq:Kmin}
\end{equation}
As long as this minimum rate is positive, 1SDI-QKD is possible. Eq.~\eqref{eq:Kmin} implies that a positive key rate is achievable for $\Pi V_{ij}<(2/e)^2$, compared to $\Pi V_{ij}<1$ for EPR-steering. Fig.~\ref{fig:Kmin} shows that in this asymmetric scheme the key rate is indeed positive between Alice and Clare over the parameter range $0.248<\zeta t<1.216$ for Alice steering Clare (i.e.~the protocol being device-independent for Alice) and $0.240<\zeta t<1.216$ for Clare steering Alice, and negative for any pair-wise coupling in the symmetric scheme.

\section{Intracavity results}
\label{sec:cav}

Having examined the asymmetric model using the solution of the Heisenberg equations of motion, which give a proof of principle, we now turn to the more experimentally realistic case where the nonlinear medium is housed inside a pumped optical cavity. With the cavity pumping field denoted by $\epsilon$, and the cavity loss rate for mode $j$ denoted by $\gamma_{j}$ (with $\gamma_{0}$ being the loss rate at the pump frequency), and again with $\kappa_{2}=0.6\kappa_{1}$, the critical pump value for the oscillation threshold is found as~\cite{Clervie}
\begin{equation}
\epsilon_{c} = \frac{\gamma_{0}\sqrt{\gamma_{1}\gamma_{2}\gamma_{3}}}{\sqrt{\kappa_{1}^{2}\gamma_{2}-\kappa_{2}^{2}\gamma_{1}}}.
\label{eq:critpump}
\end{equation}
For pump values below this, we may linearise the equations of motion and treat the system as an Ornstein-Uhlenbeck process~\cite{Ornstein}, which allows us to calculate in intracavity spectral variances as in Eq.~(11) of Pennarun \etal~\cite{Clervie}. Along with the input-output relations developed by Gardiner and Collett~\cite{Matthew}, this allows for easy calculation of output spectral correlations as a function of $\omega$ $(=\gamma_{1})$, the frequency from the cavity resonance. Although analytical results are also possible here, they become rather unwieldy, so we will present numerical results for the correlations of interest in this section. In the results presented, we use the numerical parameter values $\gamma_{0}=\gamma_{1}=\gamma_{3}=1$, $\gamma_{2}=3\gamma_{0}$, and $\kappa_{1}=0.01$. 

\begin{figure}[tbp]
\includegraphics[width=0.75\columnwidth]{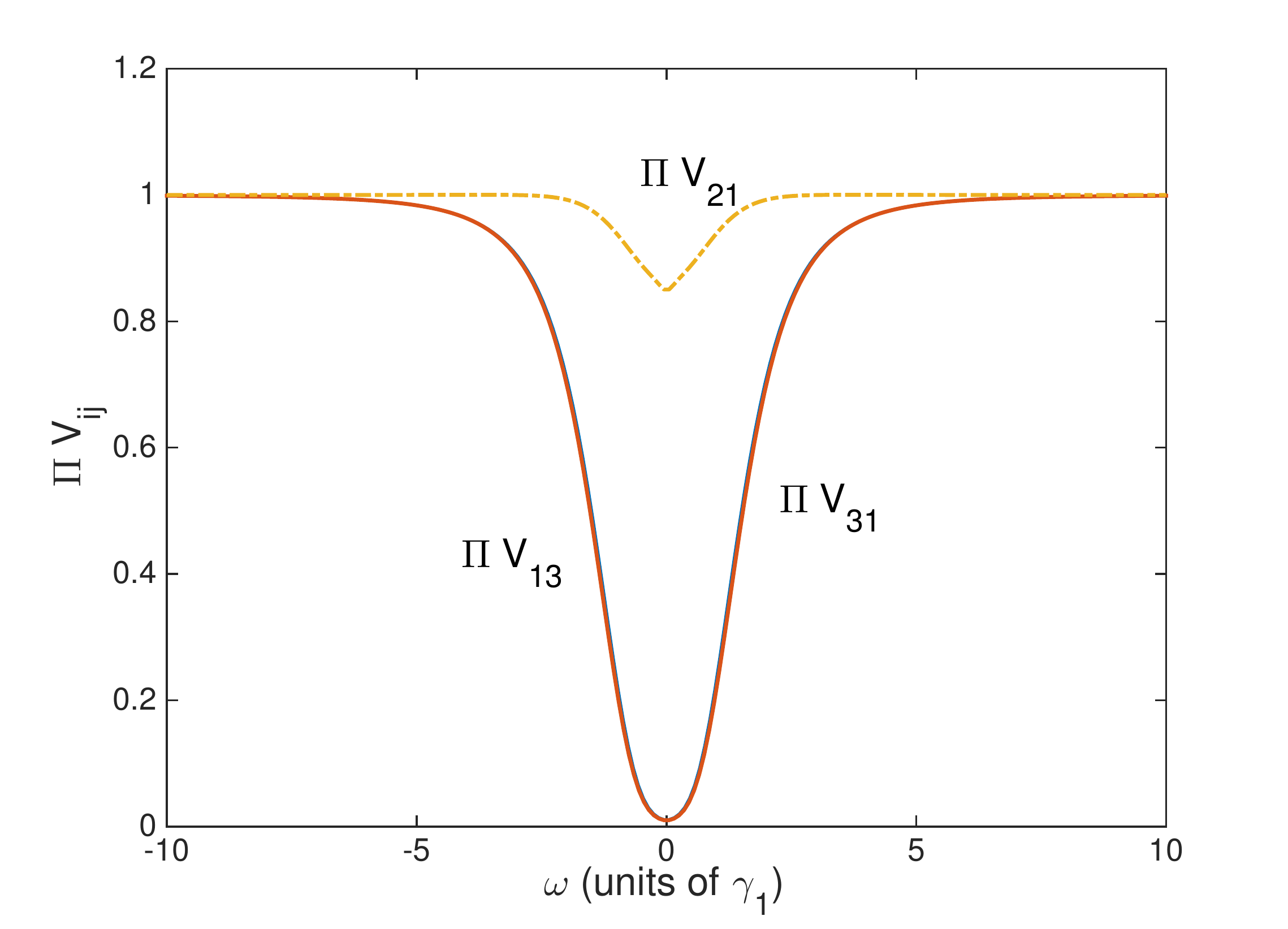}
\caption{(colour online) The output spectral EPR-steering correlations for $\epsilon=0.8\epsilon_{c}$. We see that Bob can now be steered by Alice, although the violation of the inequality is much less than those for Alice and Clare. $\omega$ represents the frequency in terms of $\gamma_{1}$ around the cavity resonance.}
\label{fig:EPRcav}
\end{figure}

\begin{figure}[tbp]
\includegraphics[width=0.75\columnwidth]{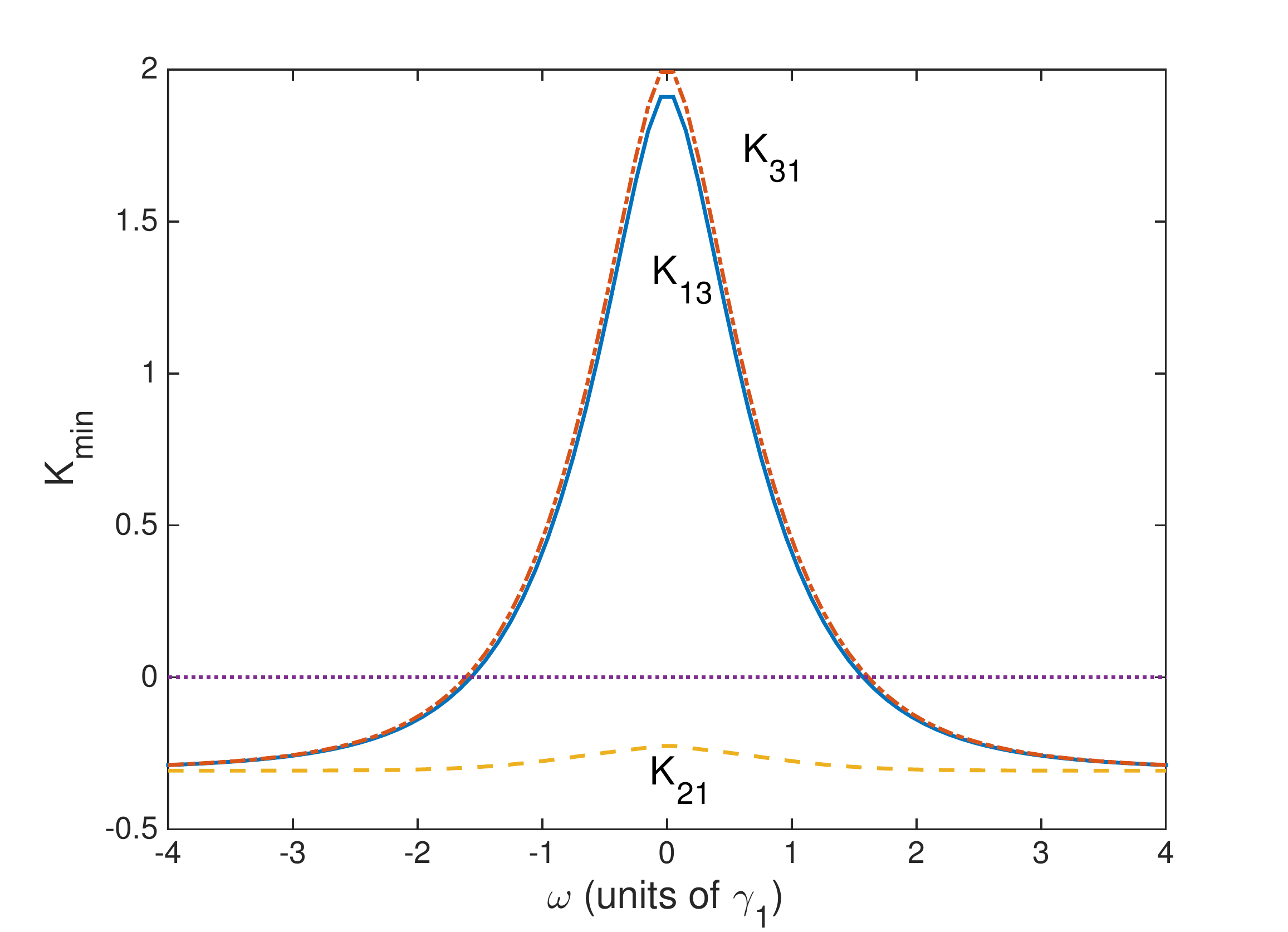}
\caption{(colour online) The spectral minimum bit rates for Alice sending a quantum key to Bob, and Alice and Clare sending to each other.  A positive value shows that 1SDI-QKD is possible. Note that the frequency axis is narrower than for Fig.~\ref{fig:EPRcav}, reflecting the fact that the bandwidth for 1SDI-QKD is less tha that for EPR-steering. The line at zero is a guide to the eye.}
\label{fig:Kmincav}
\end{figure}

\begin{figure}[tbp]
\includegraphics[width=0.75\columnwidth]{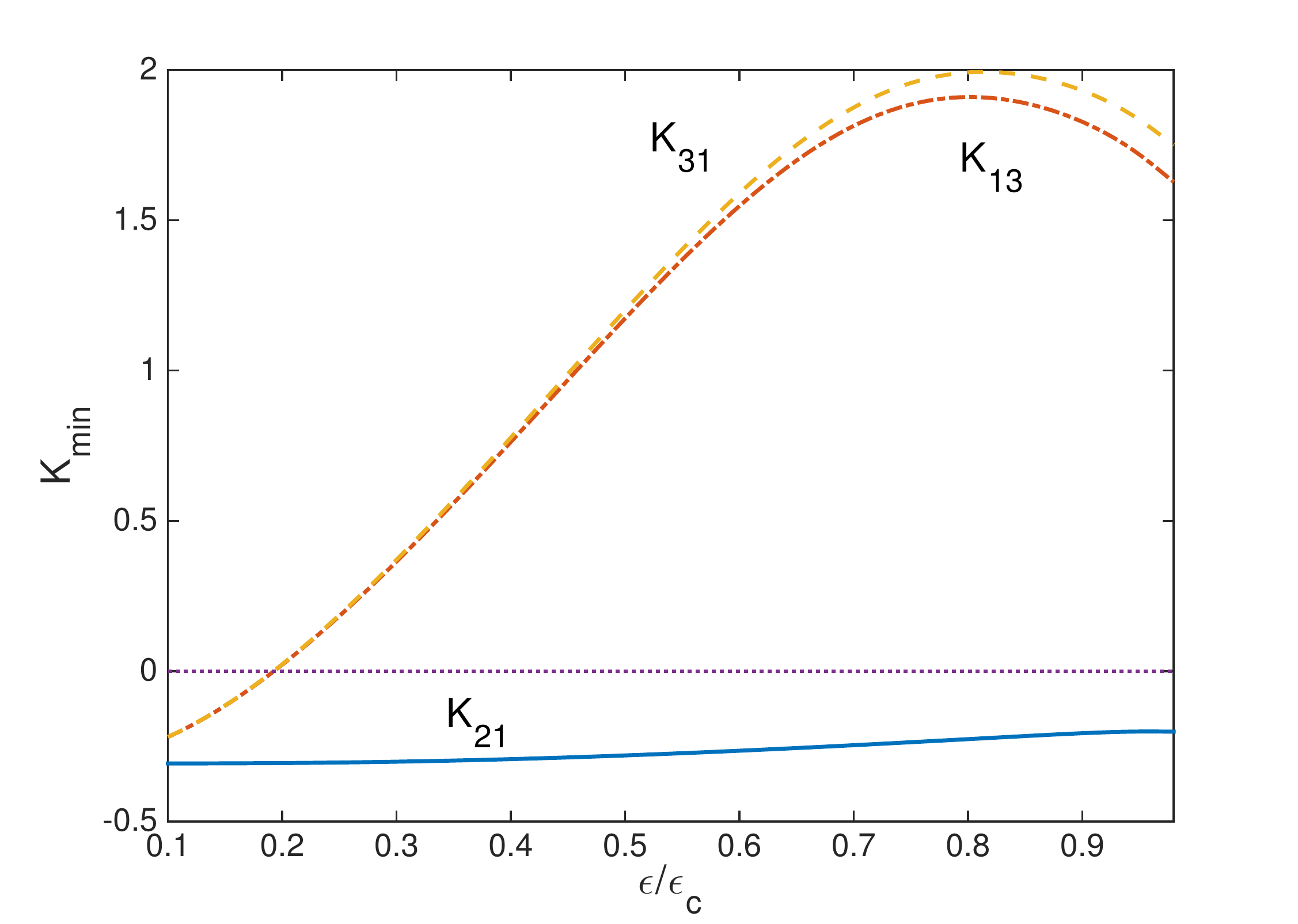}
\caption{(colour online) The minimum bit rates for modes $1$ and $3$ of the asymmetric model, with $\kappa_{2}=0.6\kappa_{1}$, and the minimum rate for any pair of the symmetric scheme. $K_{ij}$ is the minimum bit rate for mode $j$ sending a quantum key to mode $i$.  A positive value shows that 1SDI-QKD is possible. The line at zero is a guide to the eye.}
\label{fig:Kminpump}
\end{figure}

In Fig.~\ref{fig:EPRcav} we show the spectral output results for the bipartite EPR-steering correlations which show a value of less than one, for $\epsilon=0.8\epsilon_{c}$. The steering results for Alice and Clare are very close to symmetric. We also see that Alice can steer Bob to some extent, a feature which was not evident in our travelling wave model. However, this does not allow Bob to participate in 1SDI-QKD, as shown in Fig.~\ref{fig:Kmincav}, where we see that the minimum bit rate between Alice and Bob is not positive. Our predictions from the simpler model are still reliable. We also see that, as expected, the useable bandwith for 1SDI-QKD is narrower than that for EPR-steering. Since there is often excess noise in the vicinity of zero frequency, the bandwidth is an operationally important feature. Fig.~\ref{fig:Kminpump} shows the minimum bit rates as a function of $\epsilon/\epsilon_{c}$ over the range $0.1\epsilon_{c}\leq\epsilon\leq 0.98\epsilon_{c}$, with all other parameters as in the previous two figures. We again see that Alice and Clare enjoy a level of security that is not available to Bob, over this whole range.

\section{Conclusions}

We have studied the entanglement properties of an asymmetric tripartite system produced by combining downconversion with sum-frequency generation, finding that it offers an extra degree of flexibility over fully symmetric ones for QKD. With the three participants labelled as Alice, Bob, and Clare, we have shown that any pairing can share secrets using S-QKD in symmetric systems, with bipartite entanglement being available over a range of the interaction parameter. There is no bipartite 1SDI-QKD possible in these systems, with all three participants needing to be involved for the level of communication security provided by this method. On the other hand, the asymmetric system analysed here allows both bipartite S-QKD and 1SDI-QKD, with Bob only being able to participate in tripartite S-QKD. In the symmetric system, any pair of participants can also steer the remaining participant, which means that tripartite 1SDI-QKD is available as long as they work in pairs. This is not the case in the asymmetric system. In conclusion, we have shown that asymmetric Gaussian systems can offer a level of flexibility to quantum key distribution that is not available with fully symmetric systems. This may well be advantageous for some applications, for example where different levels of security are desired within a network.

{\it Acknowledgments}

This research was supported by the Australian Research Council under the Future Fellowships Program (Grant ID: FT100100515). MKO acknowledges invaluable discussions with Joel Corney.


\end{document}